\documentclass{ws-ijmpb}
\usepackage{graphicx}
\begin{document}

\markboth{Bernd Lorenz et al.}
{Superconductivity in titanium-based pnictide oxide compounds}

%
%

\title{SUPERCONDUCTIVITY IN TITANIUM-BASED PNICTIDE OXIDE COMPOUNDS}

\author{BERND LORENZ\footnote{blorenz@uh.edu}}

\address{Texas Center for Superconductivity and Department of Physics, University of Houston, \\3201 Cullen Blvd.,
Houston, Texas 77204-5002, USA}

\author{ARNOLD M. GULOY}

\address{Texas Center for Superconductivity and Department of Chemistry, University of Houston, \\3201 Cullen Blvd.,
Houston, Texas 77204-5002, USA}

\author{PAUL C. W. CHU}

\address{Texas Center for Superconductivity and Department of Physics, University of Houston, \\3201 Cullen Blvd.,
Houston, Texas 77204-5002, USA}

%

\maketitle


\begin{abstract}
Superconductivity in a novel class of layered materials, Ti-based pnictide oxides, was recently discovered. These compounds have attracted interest since they combine features of copper oxide and iron pnictide superconductors. Here the transition metal (titanium) forms two-dimensional Ti$_2$O layers (anti structure to the CuO$_2$ planes), capped by pnictogen ions (similar to Fe$_2$As$_2$ layers). The pnictide oxide compounds show a spin or charge density wave phase which coexists with superconductivity in some members of the family. Unlike the cuprates, but similar to iron pnictides, the parent compounds of pnictide oxides are metals with specific nesting properties of the Fermi surface which leads to the density wave instability. The nature of the superconductivity, coexisting with the density wave order, and the possible competition or mutual interaction between both states is one of the central questions of recent studies. This short review summarizes the current knowledge from an experimental as well as theoretical point of view and discusses some of the open questions and possible future developments.
\end{abstract}


\section{Introduction}
Superconductivity in transition metal oxides (copper oxides or cuprates), pnictides (iron arsenides), or chalcogenides (FeSe, KFe$_2$Se$_2$) have attracted increasing attention recently, not only because of the extraordinary high transition temperatures achieved in some systems, but also because of the proximity of those systems to long-range magnetic orders in form of antiferromagnetism (cuprates)\cite{chu:87,chu:11} or spin density waves (pnictides).\cite{hosono:08,ishida:09,paglione:10,stewart:11} As a consequence, "unconventional" pairing symmetries (d-wave in cuprates, s$^{+/-}$ in pnictides) mediated by spin fluctuations have been proposed to explain the superconducting properties of these systems.\cite{mazin:09} The magnetic fluctuations play a significant role\cite{lumsden:10} and evidence for a magnetic quantum critical point, hidden in the superconducting dome, was found in several pnictides.\cite{gooch:09b,maiwald:12,arsenijevic:13} Structurally, cuprates and pnictides have in common the layered structure with active superconducting layers (CuO$_2$ in cuprates and FeAs in pnictides) separated by charge reservoir blocks.

Superconductivity commonly arises through charge doping from a nonsuperconducting parent compound which, in most cuprates and pnictides, exhibits magnetic order in the ground state. Charge doping can be achieved by either varying the oxygen content (in cuprates the oxygen content can be controlled by synthesis and annealing, in pnictides oxygen can be replaced by fluorine or hydrogen) or by replacing cations with ions of different valency (for example, Sr for La in La$_2$CuO$_4$ or K for Ba in BaFe$_2$As$_2$). However, there are exemptions from the rule. LiFeAs was found to be a stoichiometric self-doped superconductor which does not need additional doping.\cite{tapp:08} Isovalent doping of phosphorous for arsenic\cite{ren:09} as well as hydrostatic pressure\cite{chu:09} have also been shown to induce superconductivity in the parent compounds of some pnictides.

In view of the exciting superconducting properties of layered cuprates and pnictides as well as the wealth of interesting physical phenomena that originate from the strong correlations between magnetism and superconductivity, various attempts have been made to search for other layered transition metal oxides/pnictides where superconductivity could be established near magnetic or other instabilities. One of the compound families with interesting magnetic and electronic properties are the Ti-based layered pnictide oxide systems of the general formula $A$Ti$_2Pn_2$O, with $A$=Na$_2$, Ba, (SrF)$_2$, etc. and $Pn$=As, Sb, Bi. The first synthesis of Na$_2$Ti$_2$As$_2$O and Na$_2$Ti$_2$Sb$_2$O by Adam and Schuster\cite{adam:90} in 1990 revealed an interesting anti-K$_2$NiF$_4$ structure, the existence of magnetic interactions in the Ti$_2$O layers resulting in a magnetic transition, but no indication of superconductivity was found.\cite{ozawa:08,johrendt:11} These materials are members of a broader pnictide oxide class of compounds the structure and properties of which have been discusses about two decades ago.\cite{brock:95} The search for superconductivity in pnictide oxides was not successful until very recently where superconductivity was discovered in BaTi$_2$Sb$_2$O and Ba$_{1-x}$Na$_x$Ti$_2$Sb$_2$O.\cite{yajima:12,doan:12} The existence of a superconducting state in pnictide oxides at low temperatures is of particular interest since superconductivity is preceded by a density wave transition at about 50 K and the possible coexistence and mutual interaction of both orders has inspired more theoretical and experimental work.

In the following sections we present a brief review of the state of the art of synthesis, structure, physical properties, and superconductivity research of Ti-based pnictide oxide compounds.

\section{Synthesis and Structure}
\subsection{Na$_2$Ti$_2Pn_2$O}
\label{SS1}
The synthesis of Na$_2$Ti$_2$$Pn_2$O ($Pn$=As, Sb) was first described by Adam and Schuster.\cite{adam:90} The solid state synthesis was achieved by mixing the precursor compounds Na$_2$O and TiAs (TiSb) at a ratio of 1:2 and annealing in Argon atmosphere for several hours at 800$^\circ$C to 900$^\circ$C. The resulting synthesis products, Na$_2$Ti$_2$As$_2$O and Na$_2$Ti$_2$Sb$_2$O, crystallize in the tetragonal structure, space group I4/mmm (No. 139). The material is unstable when exposed to air or moisture.\cite{adam:90}
Similar routes to synthesize Na$_2$Ti$_2$Sb$_2$O have been chosen by reacting stoichiometric amounts of Na$_2$O, Ti, and Sb powder in sealed Ta tubes at a temperature of 1000$^\circ$C for three days.\cite{axtell:97,ozawa:01} Liu et al. used carefully prepared Na$Pn$ powder mixed with powders of TiO$_2$ and Ti in the stoichiometric ratio to synthesize Na$_2$Ti$_2Pn_2$O through sintering at 800$^\circ$C in argon atmosphere for 50 hours.\cite{liu:09} In order to avoid contamination, all preparations had to be carried out in a glove box filled with high purity argon gas.

Single crystals of Na$_2$Ti$_2$Sb$_2$O can be grown by a NaSb flux method, as described in detail by Ozawa and Kauzlarich.\cite{ozawa:04} The polycrystalline sample of Na$_2$Ti$_2$Sb$_2$O, prepared by solid state reaction, and the NaSb flux were placed in a Ta tube enclosed in a fused silica ampoule. Crystal growth was achieved in a vertical furnace. The best crystals were obtained at a reaction temperature of 1100$^\circ$C, slow cooling at a rate of 5 to 10$^\circ$C/h, and quenching from 550$^\circ$C. This procedure yielded plate-like single crystals of several millimeters in size. Very recently, the flux method was also employed by Shi et al. to grow similarly sized single crystals of Na$_2$Ti$_2$As$_2$O and Na$_2$Ti$_2$Sb$_2$O from NaAs and NaSb flux, respectively.\cite{shi:13}

\begin{figure}[bt]
\centerline{\psfig{file=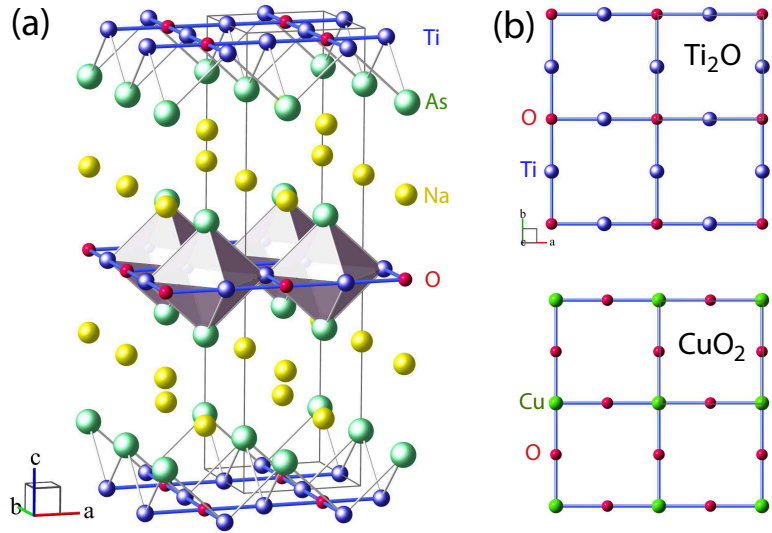,width=4in}}
\vspace*{8pt}
\caption{(a) Structure of Na$_2$Ti$_2$As$_2$O. The main building blocks are the Ti$_2$O plane capped by As ions above and below and the double layer of Na ions separating the As-Ti$_2$O-As slabs. (b) The Ti$_2$O square lattice (top) compared with the CuO$_2$ lattice (bottom) of the high-temperature superconducting cuprates.}
\label{f1}
\end{figure}

The structure type of Na$_2$Ti$_2Pn_2$O is that of Eu$_4$As$_2$O (anti-K$_2$NiF$_4$) with Na and Ti occupying the Eu (4e) and Eu (4c) sites, respectively. The tetragonal lattice of Na$_2$Ti$_2Pn_2$O ($a$=4.070 \AA, $c$=15.288 \AA$ $ for $Pn$=As) is shown in Fig. \ref{f1}a. The lattice parameters for $Pn$=Sb are slightly larger, $a$=4.144 \AA$ $ and $c$=16.561 \AA.\cite{adam:90}

The tetragonal structure of Na$_2$Ti$_2Pn_2$O consists of different layers stacked along the $c$-axis. Titanium and oxygen ions form a square lattice with oxygen occupying the nodes and titanium ions located half way between two oxygens. The $Pn$ ions are located in the center of the squares above and below the Ti$_2$O lattice. Two $Pn$ and four Ti form octahedra sharing corners only, as shown in the center of Fig. \ref{f1}a. The Na ions fill into the voids above and below the oxygen ions, forming a double layer which separates two neighboring $Pn$-Ti$_2$O-$Pn$ slabs. There are two variable parameters within the unit cell, the distance of the $Pn$ ions above and below the Ti$_2$O plane and the displacement of the Na ions from the center plane between two adjacent $Pn$-Ti$_2$O-$Pn$ slabs.

It is worth noting that the Ti$_2$O square lattice is similar to the CuO$_2$ lattice of the high-temperature superconducting cuprates, but the lattice positions of the transition metal and oxygen are interchanged (anti-K$_2$NiF$_4$ structure). Both square lattices are compared in Fig. \ref{f1}b. The similarity of the planar structures has raised the question if superconductivity could be induced into the Ti$_2$O plane through proper doping.\cite{ozawa:08}

\subsection{BaTi$_2Pn_2$O}
\label{SS2}
The successful replacement of the (Na$^+$)$_2$ double layer in Na$_2$Ti$_2$As$_2$O by a single layer of divalent Ba$^{2+}$ was first reported by Wang et al.\cite{wang:10} The synthesis resulted in a new layered pnictide oxide compound: BaTi$_2$As$_2$O. The solid state synthesis was achieved by mixing the precursors BaO, Ti, and As in a stoichiometric ratio and annealing the mixture, wrapped in tantalum foil and sealed in quartz tubes, at 850$^\circ$C for 40 hours. A polycrystalline sample of composition BaTi$_2$As$_2$O was obtained. Since Ba$^{2+}$ substituting for two Na$^+$ does not significantly change the charge state of the As-Ti$_2$O-As block, a compound with similar physical properties was obtained.

The Sb-analogue, BaTi$_2$Sb$_2$O, was synthesized by conventional solid state reaction, mixing the precursors BaO, Ti, and Sb at a ratio of 1:2:2. Two similar approaches have been reported: (i) Yajima et al.\cite{yajima:12} used tantalum foil to wrap the precursor material and sealed it in a quartz tube. The reaction was started by heating to 1000$^\circ$C for 40 hours, followed by cooling to room temperature at a rate of 25$^\circ$C/h. (ii) Doan et al.\cite{doan:12} sealed the precursors in Nb tubes under argon gas and enclosed the tubes in sealed quartz ampoules. The samples were annealed for 3 days after heating (2$^\circ$C/min) to 900$^\circ$C. Slow cooling (1$^\circ$C/min) to 200$^\circ$C and regrinding and sintering at 900$^\circ$C for another three days did ensure the homogeneity of the synthesis product.

The hole doped Ba$_{1-x}$Na$_x$Ti$_2$Sb$_2$O was successfully synthesized following a similar procedure using BaO, BaO$_2$, Na$_2$O, Na$_2$O$_2$, Ti, and Sb mixed in stoichiometric amounts as precursor materials.\cite{doan:12} Hole doping was also achieved using K or Rb to partially replace Ba. Pachmayr and Johrendt describe the synthesis of Ba$_{1-x}$K$_x$Ti$_2$Sb$_2$O from different precursor materials, depending on the potassium content.\cite{pachmayr:14} For x$<$0.5, Ba, BaO$_2$, K, Sb, and Ti were sealed in niobium tubes within silica ampoules (in argon) and slowly heated to 600$^\circ$C, held at this temperature for 15 hours, and cooled to room temperature. The reaction product was repeatedly ground, pelletized, and sintered at 900$^\circ$C for 3 days, followed by slow cooling to 200$^\circ$C to improve the chemical and structural homogeneity. For x$>$ 0.5, Ba, K, Sb, Ti, and TiO$_2$ have been used as precursors and the reaction and sintering conditions were slightly modified.\cite{pachmayr:14} Von Rohr et al. synthesized Ba$_{1-x}$Rb$_x$Ti$_2$Sb$_2$O from a stoichiometric mixture of BaO, Rb$_2$O, Ti, TiO$_2$, and Sb.\cite{rohr:13} The precursor pellets were sealed in niobium tubes and sintered at 1000$^\circ$C for 3 days. Regrinding and further sintering after the initial reaction for 24 hours at the same temperature ensured homogeneity of the final synthesis product.

Polycrystalline samples of BaTi$_2$Bi$_2$O as well as solid solutions of BaTi$_2$(As/Sb)$_2$O, BaTi$_2$(Sb/Bi)$_2$O, and BaTi$_2$(Sb/Sn)$_2$O have also been prepared by solid state reaction very recently.\cite{zhai:13,nozaki:13,yajima:13,nakano:13}

\begin{figure}[bt]
\centerline{\psfig{file=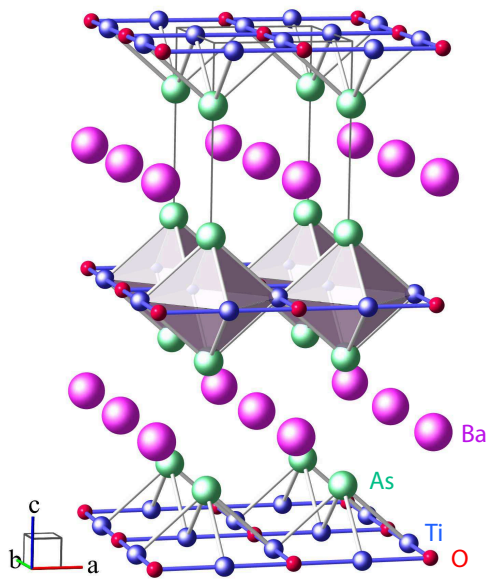,width=2.5in}}
\vspace*{8pt}
\caption{Structure of BaTi$_2$As$_2$O. Note that the $c$-axis is only half as long as in Na$_2$Ti$_2$As$_2$O.}
\label{f2}
\end{figure}

The BaTi$_2Pn_2$O compounds crystallize in the tetragonal structure with space group P4/mmm (No. 123). The structure, shown in Fig. \ref{f2}, has the identical Ti$_2$O square lattice capped above and below by $Pn$ ions as Na$_2$Ti$_2Pn_2$O (Fig. \ref{f1}a). However, the stacking of the $Pn$-Ti$_2$O-$Pn$ slabs along the $c$-axis is different, every second slab is shifted along the [1,1,0] direction so that two neighboring slabs are perfectly aligned along the $c$-direction. The Ba$^{2+}$ ions then fit exactly halfway in between two $Pn$-Ti$_2$O-$Pn$ blocks forming a square lattice on their own. Thus the double layer of Na$_2$ in Na$_2$Ti$_2Pn_2$O is replaced by a plane of Ba ions in BaTi$_2Pn_2$O. This results in a reduction of the $c$-axis and the unit cell volume by about 50\%, as shown in Fig. \ref{f2}. The symmetry leaves only one adjustable parameter in the unit cell, the distance of the $Pn$ ions from the Ti$_2$O plane.

The lattice parameters of BaTi$_2Pn_2$O increase from $Pn$=As ($a$=4.0456 \AA, $c$=7.2723 \AA)\cite{wang:10} via $Pn$=Sb ($a$=4.1196 \AA, $c$=8.0951 \AA)\cite{doan:12} to $Pn$=Bi ($a$=4.12316 \AA, $c$=8.3447 \AA)\cite{yajima:13b} with the ionic size of the pnictogen, as expected. The lattice parameters of the solid solutions BaTi$_2$(As$_{1-x}$Sb$_x$)$_2$O and BaTi$_2$(Sb$_{1-x}$Bi$_x$)$_2$O change linearly with x following Vegard's law, indicating a homogeneous mixing of ions on the pnictogen site.\cite{yajima:13}

\section{Physical properties of pnictide oxides Na$_2$Ti$_2Pn_2$O and BaTi$_2Pn_2$O}
\label{PP}
The physical properties of pnictide oxide compounds are mainly determined by the charges and spins of the titanium $d$-orbitals and the hybridized pnictogen $p$-states. Thereby, the major contribution to the Fermi surface arises from Ti $d$-states with a sharp peak of the density of states near the Fermi energy $E_F$.\cite{pickett:98} The simplified ionic picture with valencies of oxygen (2-), pnictogen (3-), Ba or Na$_2$ (2+) leaves the valence state of Ti (3+) as 3$d^1$. The one 3$d$ electron of titanium is therefore largely responsible for the transport and magnetic properties of the compounds. The existence of 3$d^1$ electrons with spin 1/2 at the Fermi energy may result in magnetic fluctuations through superexchange interactions involving the oxygen or pnictogen orbitals, as well as direct exchange between two Ti ions.

\begin{figure}[bt]
\centerline{\psfig{file=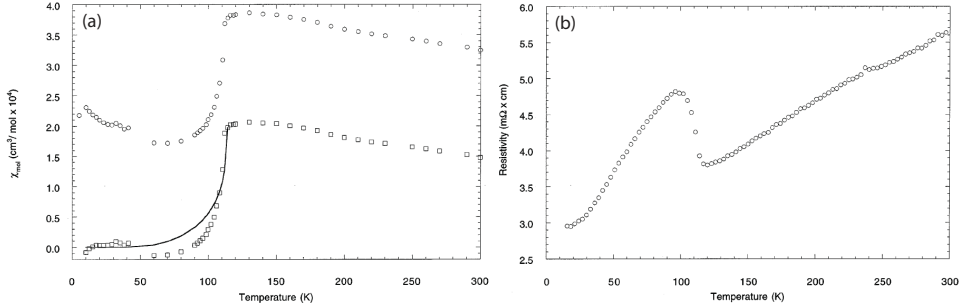,width=5in}}
\vspace*{8pt}
\caption{(a) Magnetic susceptibility and (b) resistivity of Na$_2$Ti$_2$Sb$_2$O. The lower curve in (a) was obtained after subtracting a constant (van Vleck) and a Curie-like paramagnetic contribution from the original data. Reprinted with the publisher's permission from E. A. Axtell, III, et al. J. Solid State Chem. \textbf{134}, 423 (1997).}
\label{f3}
\end{figure}

The first measurements of magnetic properties of Na$_2$Ti$_2$Sb$_2$O by Adam and Schuster have indeed revealed an anomaly of the susceptibility near a critical temperature of T$_{DW}\simeq$120 K.\cite{adam:90} The exact nature of the transition, however, is still under discussion (see Section \ref{DW}). The existence of a phase transition in Na$_2$Ti$_2$Sb$_2$O was confirmed by Axtell et al. in revealing a spin-gap like behavior of the magnetic susceptibility.\cite{axtell:97} The susceptibility shows a sharp drop below 114 K possibly indicating the opening of a spin gap in the magnetic excitation spectrum. At the transition temperature, the metallic resistivity displays a sharp increase with a subsequent decrease toward lower temperatures. The original results of Axtell et al. are shown in Fig. \ref{f3}. Other physical quantities, e.g.\ thermoelectric power and heat capacity, show clear anomalies at T$_{DW}$.\cite{liu:09} A large magnetoresistance effect of up to 40 \% was found at low temperatures and in fields up to 14 Tesla which hints for a coupling of the 3$d$ electrons to the external field.

\begin{figure}[bt]
\centerline{\psfig{file=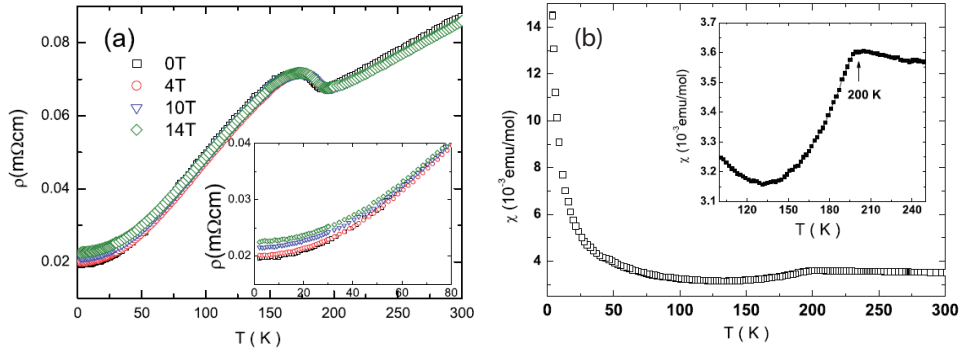,width=5in}}
\vspace*{8pt}
\caption{(a) Resistivity and (b) magnetic susceptibility of BaTi$_2$As$_2$O. The inset in (a) shows the magnetic field-effect at low temperatures on an enlarged scale. Reprinted with the publisher's permission from X. F. Wang, et al. J. Phys.: Condens. Matter \textbf{22}, 075702 (2010).}
\label{f4}
\end{figure}

The possible origin of the instability in Na$_2$Ti$_2$Sb$_2$O was addressed by Pickett\cite{pickett:98} based on first-principles calculations of the electronic structure (see also Section \ref{FPC}). The metallic character of Na$_2$Ti$_2$Sb$_2$O was explained by a strong mixing of the Ti $d$-states with the $p$-states of Sb and O and the direct $d$-$d$ overlap. The Fermi surface, dominated by the delocalized Ti 3$d$ states, exhibits a nesting property which apparently leads to the observed instability at T$_{DW}$. The low-temperature state could be a spin-density wave (SDW) or a charge-density wave (CDW) state, but the calculations did not allow to decide which order may be realized in Na$_2$Ti$_2$Sb$_2$O.

A similar drop of the magnetic susceptibility, although not as sharp, was found in Na$_2$Ti$_2$As$_2$O at about 320 K,\cite{ozawa:01} but a clear anomaly in the resistivity could not be resolved.\cite{liu:09} The overall transport characteristics of Na$_2$Ti$_2$As$_2$O shows a negative temperature coefficient, in contrast to the metallic behavior of Na$_2$Ti$_2$Sb$_2$O. The significantly higher T$_{DW}$ of Na$_2$Ti$_2$As$_2$O shows that the pnictogen ion has a strong influence on the stability range of the density wave phase.

BaTi$_2$As$_2$O, first synthesized by Wang et al.\cite{wang:10} in 2010, exhibits the density wave transition at T$_{DW}$=200 K, significantly lower than Na$_2$Ti$_2$As$_2$O. A clear drop of the magnetic susceptibility and a sharp increase of the resistivity, followed by a metallic-like decrease to lower temperature, as well as a pronounced peak of the heat capacity signal the onset of the density wave order. Resistivity and magnetic susceptibility data from Ref. [26] are shown in Fig. \ref{f4}. Attempts to induce superconductivity by charge doping into the Ti$_2$O plane through substitution of barium by potassium or sodium as well as intercalation of lithium have not been successful. However, the density wave transition was shifted to lower temperature in the intercalated Li$_x$BaTi$_2$As$_2$O.

The question whether or not superconductivity can be induced in the Ti-based layered pnictide oxide compounds could be related to the stability of the density wave phase. If the spin or charge ordered states compete with superconductivity, the suppression of the SDW/CDW phase should be in favor of a superconducting state. Comparing T$_{DW}$ of Na$_2$Ti$_2Pn_2$O for $Pn$=As and $Pn$=Sb, it becomes obvious that the replacement of As by Sb reduces the critical temperature significantly, namely from 320 K (As) to 114 K (Sb). On the other hand, the $T_{DW}$ of BaTi$_2$As$_2$O (200 K) is also much smaller than that of Na$_2$Ti$_2$As$_2$O (320 K). The replacement of As by Sb or Bi in BaTi$_2$As$_2$O could further reduce or completely suppress the density wave state.

\begin{figure}[bt]
\centerline{\psfig{file=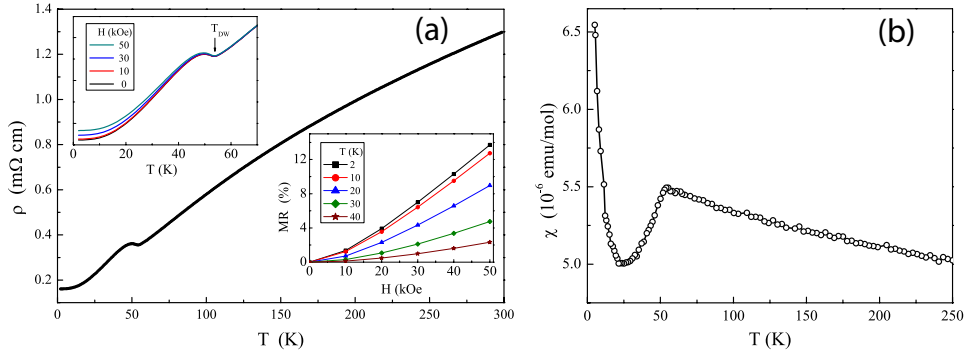,width=5in}}
\vspace*{8pt}
\caption{(a) Resistivity and (b) magnetic susceptibility of BaTi$_2$Sb$_2$O. The upper inset in (a) shows the magnetic field-effect at low temperatures on an enlarged scale. The lower inset displays the magnetoresistance at several constant temperatures.}
\label{f5}
\end{figure}

The physical properties of BaTi$_2$Sb$_2$O, first synthesized in 2012,\cite{yajima:12,doan:12} indeed show the decrease of T$_{DW}$ to 54 K. Resistivity and magnetic susceptibility data for BaTi$_2$Sb$_2$O are shown in Fig. \ref{f5}. The resistivity exhibits a hump-like feature with a sharp increase at T$_{DW}$ and a large magnetoresistance effect at lower temperatures, as shown in the upper inset of Fig. \ref{f5}a, similar to the results for BaTi$_2$As$_2$O (Fig. \ref{f4}a). The magnitude of the magnetoresistance at 50 kOe exceeds the one observed in the As analogue significantly.\cite{wang:10} The percentage increase, $(\rho(H)-\rho(0))/\rho(0)$, at 2 K and 50 kOe reaches 14 \%. The density wave transition in BaTi$_2$Sb$_2$O is also revealed by a sharp drop of the magnetic susceptibility (Fig. \ref{f5}b).

\begin{figure}[bt]
\centerline{\psfig{file=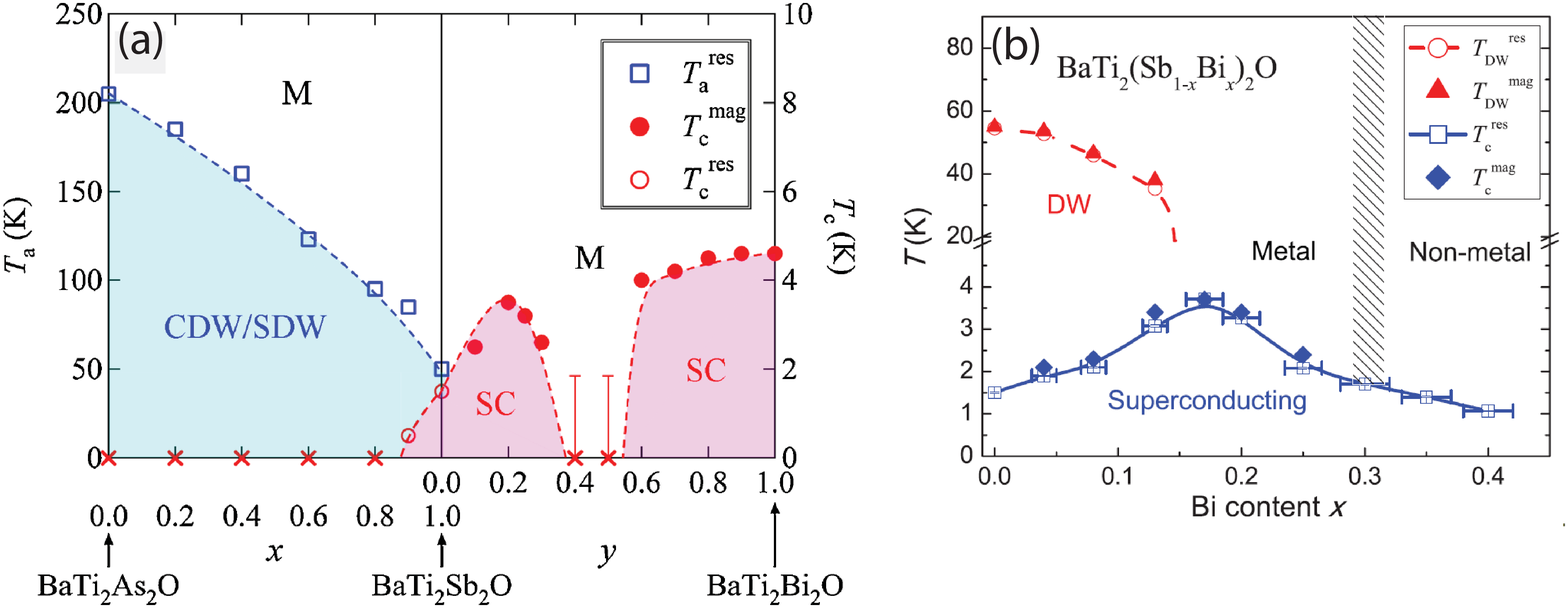,width=5in}}
\vspace*{8pt}
\caption{(a) Phase diagram of BaTi$_2$(As$_{1-x}$Sb$_x$)$_2$O (left half) and BaTi$_2$(Sb$_{1-x}$Bi$_x$)$_2$O (right half). Reprinted with the publisher's permission from T. Yajima et al., J. Phys. Soc. Jpn. \textbf{82}, 033705 (2013). (b) Phase diagram of BaTi$_2$(Sb$_{1-x}$Bi$_x$)$_2$O in the range of low substitution, x$<$0.4. Note the further reduction of T$_{DW}$ upon Bi substitution. Reprinted with the publisher's permission from H.-F. Zhai et al., Phys. Rev. B \textbf{87}, 100502(R) (2013).}
\label{f6}
\end{figure}

It is obvious that the critical temperature of the density wave phase is significantly lower in BaTi$_2Pn_2$O and it is further reduced by replacing arsenic with antimony. It should be noted that this phase is completely absent at all temperatures in BaTi$_2$Bi$_2$O.\cite{yajima:13} A systematic decrease of T$_{DW}$ in BaTi$_2$(As$_{1-x}$Sb$_x$)$_2$O from 200 K to 54 K and in BaTi$_2$(Sb$_{1-x}$Bi$_x$)$_2$O from 54 K to zero temperature prove the suppression of the DW phase with increasing (average) size of the pnictogen ion.\cite{zhai:13,yajima:13} The phase diagrams for BaTi$_2$(As$_{1-x}$Sb$_x$)$_2$O and BaTi$_2$(Sb$_{1-x}$Bi$_x$)$_2$O are shown in Fig. \ref{f6}a and Fig. \ref{f6}b, respectively. The extrapolation of T$_{DW}$ to zero temperature yields a critical value of the Bi content between x=0.15 and x=0.2.

\section{Superconductivity in pnictide oxide compopunds}
\label{SC}
\subsection{Hole-doped BaTi$_2$Sb$_2$O}
\label{SCH}
\begin{figure}[bt]
\centerline{\psfig{file=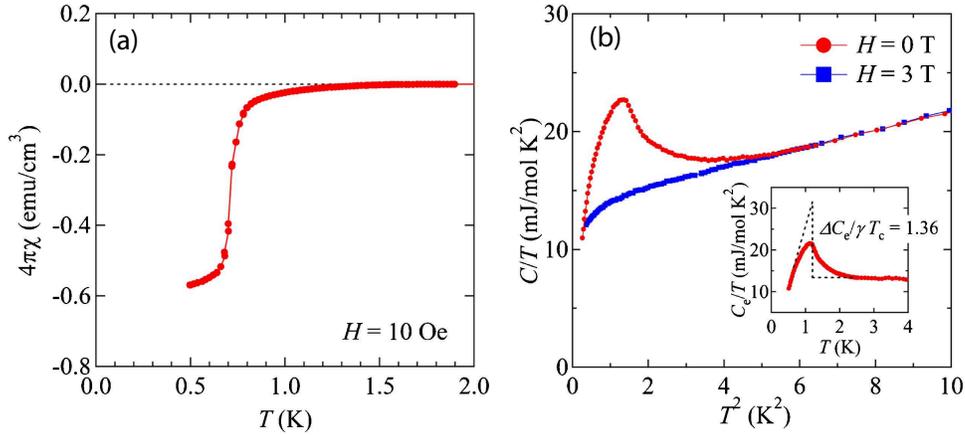,width=5in}}
\vspace*{8pt}
\caption{(a) Superconducting shielding signal and (b) heat capacity of BaTi$_2$Sb$_2$O. Reprinted with the publisher's permission from T. Yajima et al., J. Phys. Soc. Jpn. \textbf{81}, 103706 (2012).}
\label{f7}
\end{figure}

Superconductivity in the pnictide oxides was discovered simultaneously in BaTi$_2$Sb$_2$O with a critical temperature of T$_c$=1.2 K\cite{yajima:12} and in the sodium substituted Ba$_{1-x}$Na$_x$Ti$_2$Sb$_2$O with T$_c$ up to 5.5 K.\cite{doan:12} The bulk nature of the superconducting state in BaTi$_2$Sb$_2$O is proven by a significant magnetic shielding signal and a pronounced peak of the heat capacity, as shown in Figs. \ref{f7}a and \ref{f7}b, respectively. The drop of the resistivity to zero confirms the existence of a superconducting phase below 1.2 K.

\begin{figure}[bt]
\centerline{\psfig{file=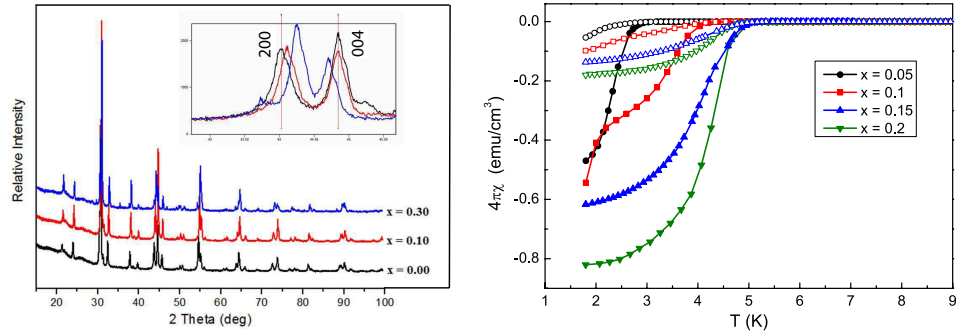,width=5in}}
\vspace*{8pt}
\caption{(a) Selected X-ray spectra of hole-doped Ba$_{1-x}$Na$_x$Ti$_2$Sb$_2$O. (b) Low-temperature magnetic susceptibility of Ba$_{1-x}$Na$_x$Ti$_2$Sb$_2$O measured in 10 Oe applied field. Bold symbols: Shielding signal (zero field cooled), Open symbols: Meissner signal (field cooled).}
\label{f8}
\end{figure}

Replacing Ba by Na removes electrons from the active layer, i.e. Na substitution results in hole doping of the Ti$_2$O plane. The synthesis of Ba$_{1-x}$Na$_x$Ti$_2$Sb$_2$O was successful for x$\leq$0.35. X-ray spectra for x=0, 0.1, 0.3 are shown in Fig. \ref{f8}a. The systematic shift of the [200] and [004] peak positions, shown in the inset, indicate that the Na ions dope into the Ba positions. The solubility limit for Na in the P4/mmm structure was found at about 35\%. Magnetic susceptibility data in Fig. \ref{f8}b prove the existence of superconductivity in Ba$_{1-x}$Na$_x$Ti$_2$Sb$_2$O with T$_c$ systematically increasing with the Na content up to x=0.2. The large shielding signal and a significant Meissner effect indicate the bulk nature of the superconducting phase. Similar susceptibility data for Ba$_{1-x}$Na$_x$Ti$_2$Sb$_2$O have been reported recently and confirm the results of Fig. \ref{f8}b with regard to the magnitude of magnetic shielding as well as the critical temperatures.\cite{rohr:13b}

\begin{figure}[bt]
\centerline{\psfig{file=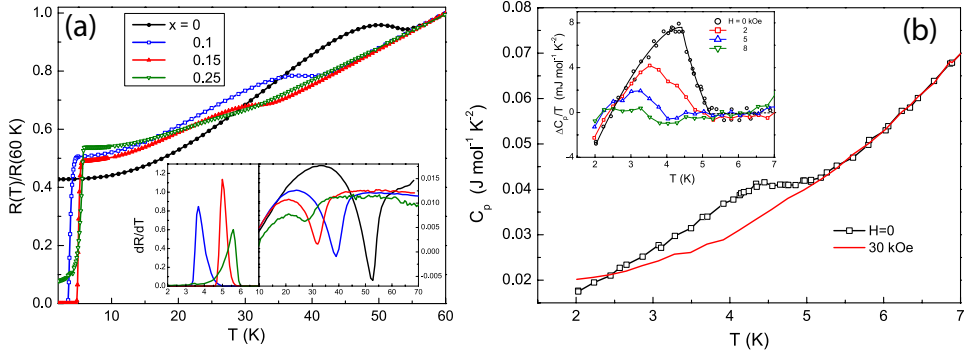,width=5in}}
\vspace*{8pt}
\caption{(a) Resistance of Ba$_{1-x}$Na$_x$Ti$_2$Sb$_2$O for various Na concentrations x. The inset shows the derivative used to define T$_c$ (peak, left panel) and T$_{DW}$ (dip, right panel). (b) Heat capacity of Ba$_{0.85}$Na$_{0.15}$Ti$_2$Sb$_2$O.}
\label{f9}
\end{figure}

\begin{figure}[bt]
\centerline{\psfig{file=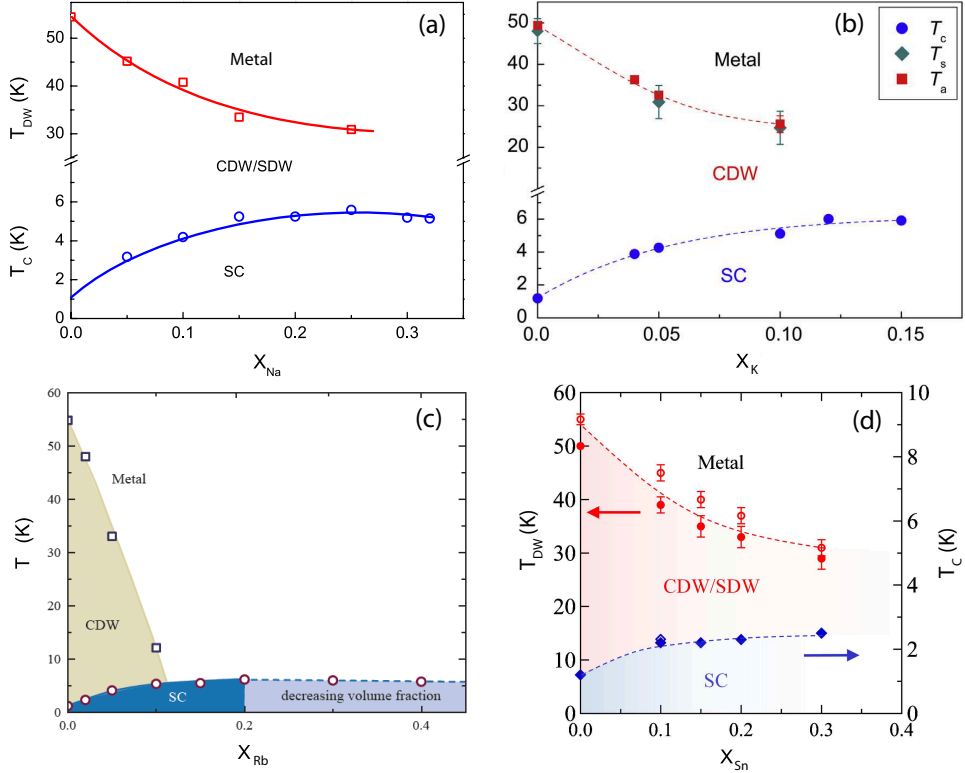,width=5in}}
\vspace*{8pt}
\caption{Phase diagrams of hole-doped BaTi$_2$Sb$_2$O. (a) Ba$_{1-x}$Na$_x$Ti$_2$Sb$_2$O (Ref. [20]), (b) Ba$_{1-x}$K$_x$Ti$_2$Sb$_2$O, (Reprinted with the publisher's permission from U. Pachmayr and D. Johrendt, Solid State Sciences \textbf{28}, 31 (2014)), (c) Ba$_{1-x}$Rb$_x$Ti$_2$Sb$_2$O, (Reprinted with the publisher's permission from F. von Rohr et al., Phys. Rev. B \textbf{89}, 094505 (2014)), (d) BaTi$_2$(Sb$_{1-x}$Sn$_x$)$_2$O, (Reprinted with the publisher's permission from K. Nakano et al., J. Phys. Soc. Jpn. \textbf{82}, 074707 (2013))}
\label{f10}
\end{figure}

The increase of T$_c$ with the sodium content shows that the stability of the superconducting state is controlled by the charge carrier density, similar to copper and iron based high-temperature superconductors. More interestingly, the characteristic signature of the DW transition, a hump-like anomaly in the resistivity curves, is still observed for sodium contents up to 25\%, as shown in Fig. \ref{f9}a. The sizable peak of the heat capacity and it's characteristic dependence on magnetic fields, shown in Fig. \ref{f9}b for x=0.15, proves the bulk nature of superconductivity in the sodium doped compound. The superconducting phase coexists with the density wave phase at low temperatures. The critical temperature T$_{DW}$, as determined from the dip in $dR/dT$ (right inset in Fig. \ref{f9}a), seems to level off with increasing x at about 30 K. This is in contrast to the spin density wave phase of the iron pnitide superconductors where the SDW transition is sharply suppressed upon charge doping and it extrapolates to zero at a critical doping, revealing a (hidden) quantum critical point. For Na concentrations above 25 \%, the DW anomaly could not be resolved any more in resistivity measurements. The phase diagram of Ba$_{1-x}$Na$_x$Ti$_2$Sb$_2$O is shown in Fig. \ref{f10}a for $x$ values up to the Na solubility limit.

Hole doping by alkali metal substitution for barium was also achieved using potassium and rubidium. The increase of T$_c$ and the decrease of T$_{DW}$, reported in Ba$_{1-x}$K$_x$Ti$_2$Sb$_2$O,\cite{pachmayr:14} are shown in the phase diagram of Fig. \ref{f10}b. The maximum of T$_c\simeq$6.1 K was obtained at x$_K$=0.12 in Ba$_{1-x}$K$_x$Ti$_2$Sb$_2$O with a quickly decreasing superconducting shielding signal at higher doping concentrations. The decrease of T$_{DW}$ with x$_K$ with a concave curvature and an apparent saturation above x$_K$=0.1 mimics the phase diagram of Ba$_{1-x}$Na$_x$Ti$_2$Sb$_2$O.

Doping rubidium into the barium site increases T$_c$ of Ba$_{1-x}$Rb$_x$Ti$_2$Sb$_2$O up to 5.4 K at optimal doping of x=0.2, similar to Ba$_{1-x}$Na$_x$Ti$_2$Sb$_2$O.\cite{rohr:13} The critical temperature of the DW phase, however, appeared to decrease faster and T$_{DW}$ was extrapolated to zero at a critical doping of x=0.12. This behavior seems to be different from other hole-doped BaTi$_2$Sb$_2$O compounds, however, one should take into account that the definition of T$_{DW}$ from resistivity data for larger x values becomes less certain due to the disappearance of the characteristic hump-like anomaly.

Whereas alkali metal ions like Na, K, Rb replace the Ba ions outside the Sb-Ti$_2$O-Sb layer with a minimal structural impact on the active superconducting block, hole doping can also be achieved by substituting tin for antimony in BaTi$_2$(Sb$_{1-x}$Sn$_x$)$_2$O.\cite{nakano:13} The doping-induced increase of T$_c$, however, was significantly smaller with a maximum of 2.5 K at x=0.3. The phase diagram and the decrease of T$_{DW}$ with the concave curvature is very similar to the cases where Ba was replaced by Na and K. The smaller T$_c$ in BaTi$_2$(Sb$_{1-x}$Sn$_x$)$_2$O may be due to disorder effects introduced into the active Sb-Ti$_2$O-Sb block. The partially covalent bond between the Ti and Sb ions are perturbed by Sn substitution which gives rise to chemical disorder affecting the hybridization of titanium and pnictogen states at the Fermi energy. The covalent character of the Ti-Sb bond was also suggested to reduce the effective, tin doping-induced charge transfer to the Ti$_2$O plane.\cite{nakano:13}

The existing gross similarities, but also the subtle differences, in the physical properties and the superconducting/density wave phase diagrams of hole-doped pnictide oxides raise the question whether or not changes of the structural parameters induced by the ionic substitutions are essential, besides the obvious doping of holes into the active layer. The effects of potassium, sodium, and rubidium doping on the lattice constants of Ba$_{1-x}$A$_x$Ti$_2$Sb$_2$O (A=Na, K, Rb) are qualitatively similar with $a$ decreasing and $c$ increasing with x, however, the overall volume change is different. In Ba$_{1-x}$Na$_x$Ti$_2$Sb$_2$O the unit cell volume clearly decreases whereas Ba$_{1-x}$K$_x$Ti$_2$Sb$_2$O shows a slight but continuous increase with x.\cite{pachmayr:14} The volume of Ba$_{1-x}$Rb$_x$Ti$_2$Sb$_2$O also decreases, but the rate of decrease is much smaller than in the Na-doped compound.\cite{rohr:13} The effect of tin doping is different, both lattice parameters $a$ and $c$ decrease with x, resulting in a volume decrease comparable with that of Ba$_{1-x}$Na$_x$Ti$_2$Sb$_2$O.\cite{nakano:13}

Despite the different responses of the unit cell parameters to doping, the stability of the superconducting and DW phases and the resulting phase diagrams are very similar, as shown in Fig. \ref{f10}. This indicates that the charge transfer to or from the active layer is the most relevant parameter determining the properties of hole-doped BaTi$_2$Sb$_2$O.

\subsection{Isovalent substitution of Sb by As and Bi}
\label{SCI}
Isovalent doping of BaTi$_2$Sb$_2$O with Bi was reported to increase the superconducting critical temperature of BaTi$_2$(Sb$_{1-x}$Bi$_x$)$_2$O up to 3.7 K at the optimal x=0.17. At higher x values (up to x=0.4), T$_c$ decreased again and the temperature coefficient of the resistivity changed sign, indicating a metal-nonmetal transition near x=0.3.\cite{zhai:13} The DW phase was further suppressed with T$_{DW}$ decreasing to about 40 K at x=0.13 and the characteristic anomaly in resistivity and magnetization could not be discerned any more at x=0.17. The phase diagram derived from resistivity and magnetization measurements is shown in Fig. \ref{f6}b. The decrease of T$_{DW}$ follows the trend with increasing ionic radius of the pnictogen, starting from BaTi$_2$As$_2$O (T$_{DW}$=200 K) to BaTi$_2$Sb$_2$O (T$_{DW}$=54 K) and finally to BaTi$_2$(Sb$_{0.87}$Bi$_{0.13}$)$_2$O (T$_{DW}$=40 K).

It is important to consider any possible impurity phase which might be superconducting in Bi containing compounds. It is known, for example, that BaBi$_3$ is superconducting at 5.7 K.\cite{matthias:52} Pure bismuth, not superconducting in the crystalline state, can become superconducting in its amorphous form, with T$_c$ above 6 K.\cite{shier:66} In BaTi$_2$(Sb$_{1-x}$Bi$_x$)$_2$O, the onset of a resistance drop as well as a small diamagnetic signal at higher temperatures (near 6 K) was indeed observed and attributed to Bi impurities, also resolved in X-ray spectra.\cite{zhai:13}

Yajima et al. have synthesized the complete series of solid solutions of BaTi$_2$(Sb$_{1-x}$As$_x$)$_2$O and BaTi$_2$(Sb$_{1-x}$Bi$_x$)$_2$O for 0$\leq$x$\leq$1.\cite{yajima:13} The smooth increase of both lattice parameters with x, following Vegard's law over the whole concentration range, shows that the compounds form solid solutions. The complete phase diagram is shown in Fig. \ref{f6}a. T$_{DW}$ decreases smoothly from 200 K to 54 K in BaTi$_2$(Sb$_{1-x}$As$_x$)$_2$O, as expected. The superconducting T$_c$, in BaTi$_2$Sb$_2$O at 1.2 K, decreases to zero when a small amount of As is substituted for Sb. The dome-shaped stability region of the superconducting phase for Bi doping up to x=0.4 was also observed, similar to the report by Zhai et al.\cite{zhai:13} However, at higher Bi doping (x$\geq$0.6), a reentrant superconducting phase was found.\cite{yajima:13} The T$_c$ of this phase reaches a maximum of 4.6 K for x=1 (BaTi$_2$Bi$_2$O). Based on magnetic (zero field cooled) measurements, showing a relatively large shielding signal, the authors suggested that bulk superconductivity in this phase is intrinsic.

\subsection{Superconducting properties of pnictide oxides}
\label{SCP}
In view of the unconventional superconducting properties and pairing symmetries of the layered cuprates (d-wave pairing) and iron arsenides (s$^\pm$ pairing, two-gap superconductivity), the physical nature and the details of the superconducting state and the symmetry of the order parameter in pnictide oxides are of essential interest. Since in BaTi$_2$Sb$_2$O and it's hole doped variants superconductivity coexists with the density wave order, the possible competition or mutual interaction of both states may result in unconventional pairing mechanisms. Two possible scenarios have been discussed theoretically in recent first-principles calculations of BaTi$_2$Sb$_2$O and it's hole-doped analogue.\cite{singh:12,subedi:13} Interestingly, the possible symmetry of the superconducting gap depends on the nature of the density wave phase, spin (SDW) or charge density wave (CDW). A magnetic instability, caused by Fermi surface nesting, would result in the formation of a spin density wave and allow for spin-fluctuation mediated superconductivity and the possibility of a sign-changing s-wave superconducting state.\cite{singh:12} Contrary, when the Fermi surface nesting results in a charge density wave, electron-phonon mediated superconductivity with the possibility of multiband character was suggested based on first-principle calculations of the phonon dispersions and the electron-phonon coupling.\cite{subedi:13} Details of the first-principles calculations will be discussed in Section \ref{FPC}.

\subsubsection{Heat capacity}
\label{CP}
\begin{figure}[bt]
\centerline{\psfig{file=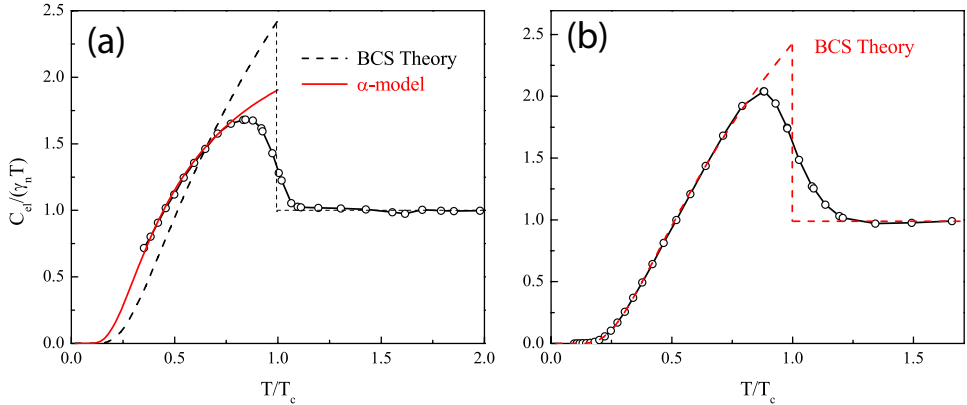,width=5in}}
\vspace*{8pt}
\caption{Electronic heat capacity of (a) BaTi$_2$Sb$_2$O and (b) Ba$_{0.85}$Na$_{0.15}$Ti$_2$Sb$_2$O. The dashed lines show the fit to the BCS theory. The C$_p$ data of BaTi$_2$Sb$_2$O are better described by the single-gap $\alpha$ model, as discussed in the text.}
\label{f11}
\end{figure}

The heat capacity of the superconducting state is a sensitive quantity, reflecting unconventional pairing symmetries or multigap superconductivity. Data available so far will be discussed in the following section. Because of the low superconducting transition temperatures (1.2 K to 5.5 K), meaningful heat capacity (C$_p$) measurements have to be extended to temperatures below 1 K. A systematic study of the superconducting C$_p$ to $^3$He temperatures has been conducted for the hole-doped Ba$_{1-x}$Na$_x$Ti$_2$Sb$_2$O.\cite{gooch:13} The results for x=0 and x=0.15 are shown in Fig. \ref{f11}. The heat capacity of both compounds can be fit to a single-gap weak-coupling model. For 15\% Na doping, the data fit nearly perfectly to the BCS model, as shown by the dashed line in Fig. \ref{f11}b. The data for the undoped BaTi$_2$Sb$_2$O, however, deviate significantly from the BCS function (Fig. \ref{f11}a). The best fit is provided by the $\alpha$-model\cite{padamsee:73} which varies the zero-temperature superconducting gap value $\Delta(0)$ while maintaining the same relative temperature dependence $\Delta(T)/\Delta(0)$ according to the BCS model. The parameter $\alpha$ is defined as $\alpha=\Delta(0)/k_BT_c$ and the best fit for BaTi$_2$Sb$_2$O is obtained for $\alpha\approx$1.4 (note that $\alpha$=1.764 in the BCS theory). The normal state Sommerfeld constant, $\gamma_n$, is obtained as 10.9 mJ/(mol K$^2$) and 13 mJ/(mol K$^2$) for x=0 and x=0.15, respectively, in good agreement with calculated values.\cite{singh:12}

The heat capacity data of BaTi$_2$Sb$_2$O, shown in Fig. \ref{f11}a, are consistent with data reported by Yajima et al.\cite{yajima:12}, however, in the latter case the large width of the superconducting transition made it difficult to fit C$_p$(T) to a more detailed model (see Fig. \ref{f7}b).

Results of heat capacity measurements of other hole-doped BaTi$_2$Sb$_2$O compounds are consistent with the above conclusions. In Ba$_{0.8}$Rb$_{0.2}$Ti$_2$Sb$_2$O, the heat capacity jump at T$_c$=5.4 K, $\Delta C_p/\gamma_n T_c$=1.57, is compatible with the BCS model and the normal state $\gamma_n$=14 mJ/(mol K$^2$) is nearly the same as the value for Ba$_{0.85}$Na$_{0.15}$Ti$_2$Sb$_2$O discussed above.\cite{rohr:13} A more detailed fit to the BCS or $\alpha$-models seems not feasible since the data are limited to T$>$2 K. Heat capacity data of BaTi$_2$(Sb$_{0.9}$Sn$_{0.1}$)$_2$O have also been found to be consistent with a single-gap BCS model with $\Delta C_p/\gamma_n T_c$=1.39 at T$_c$=2.3 K and $\gamma_n$=17.1 mJ/(mol K$^2$).\cite{nakano:13}

All heat capacity data published so far are consistent with a weak-coupling single-gap BCS description, although the unconventional, spin-fluctuation mediated pairing mechanism\cite{singh:12} cannot be completely ruled out. Furthermore, the data do not show any signature of multiple superconducting gaps, as suggested by Subedi.\cite{subedi:13} Further evidence for conventional weak-coupling superconductivity in pnictide oxides was derived from muon spin rotation measurements (London penetration depth) of Ba$_{1-x}$Na$_x$Ti$_2$Sb$_2$O\cite{rohr:13b} and BaTi$_2$(As$_{1-x}$Sb$_x$)$_2$O,\cite{nozaki:13} with no clear indication of multiple superconducting gaps. Support for an ordinary s-wave superconducting state in BaTi$_2$Sb$_2$O was also derived from $^{121/123}$Sb nuclear magnetic and quadrupole resonance experiments revealing the existence of a coherence peak at T$_c$ in the 1/T$_1$ vs. temperature data.\cite{kitagawa:13}

\subsubsection{Upper and lower critical fields}
\label{CF}
\begin{figure}[bt]
\centerline{\psfig{file=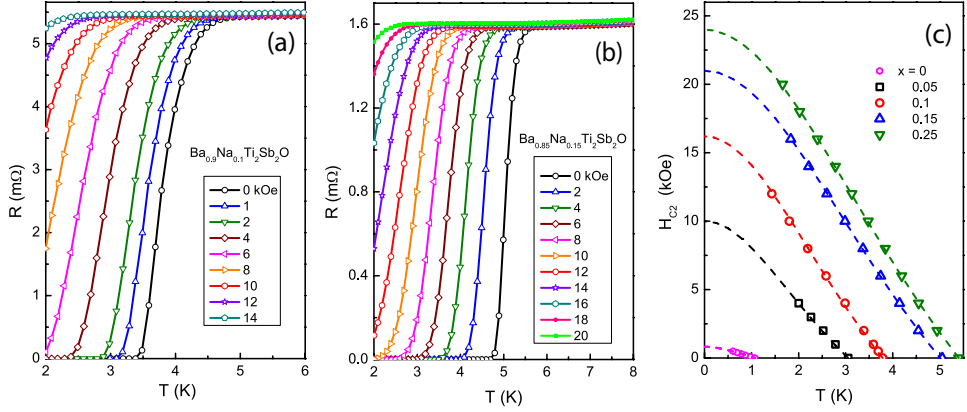,width=5in}}
\vspace*{8pt}
\caption{Resistivity of Ba$_{1-x}$Na$_x$Ti$_2$Sb$_2$O in magnetic fields. (a) x=0.1  (b) x=0.15  (c) Upper critical fields H$_{c2}$ as derived from resistivity data.}
\label{f12}
\end{figure}

The upper critical field, H$_{c2}$, can be determined from field-dependent resistivity measurements. An example is shown in Fig. \ref{f12} for the system Ba$_{1-x}$Na$_x$Ti$_2$Sb$_2$O. The extrapolation to zero temperature (dashed lines in Fig. \ref{f12}c) yields H$_{c2}$ values between 0.8 kOe (x=0) and 24 kOe (x=0.25). These numbers can be used to estimate the (average) coherence length $\xi(0)$ which decreases from 640 {\AA} for x=0 to 120 {\AA} for x=0.25. These data are in excellent agreement with other hole-doped BaTi$_2$Sb$_2$O compounds.\cite{rohr:13}

The lower critical field H$_{c1}$, which is related to the London penetration depth $\lambda$, was measured for the optimally doped Ba$_{1-x}$Rb$_x$Ti$_2$Sb$_2$O.\cite{rohr:13} With the value of H$_{c1}\approx$ 38 Oe and the upper critical field H$_{c2}$=23 kOe, the Ginzburg-Landau parameter can be estimated as $\kappa\approx$35, revealing that pnictide oxides are strongly type-II superconductors. Accordingly, the magnetic penetration depth for this compound is $\lambda=\kappa\xi\approx$4200 {\AA}.\cite{rohr:13}

\subsubsection{Pressure effects on superconductivity}
\label{PE}
The effect of hydrostatic pressure on the superconducting state was studied extensively for conventional as well as unconventional superconductors.\cite{lorenz:05,chu:09} The control of fundamental parameters, like the phonon energy, electron-phonon coupling strength, charge transfer and chemical bonds, etc. by external pressure result in complex responses of the superconducting state to the lattice compression. On the other hand, a systematic study of the stability of the superconducting state with respect to pressure can also illuminate the microscopic mechanisms and fundamental interactions leading to superconductivity.

\begin{figure}[bt]
\centerline{\psfig{file=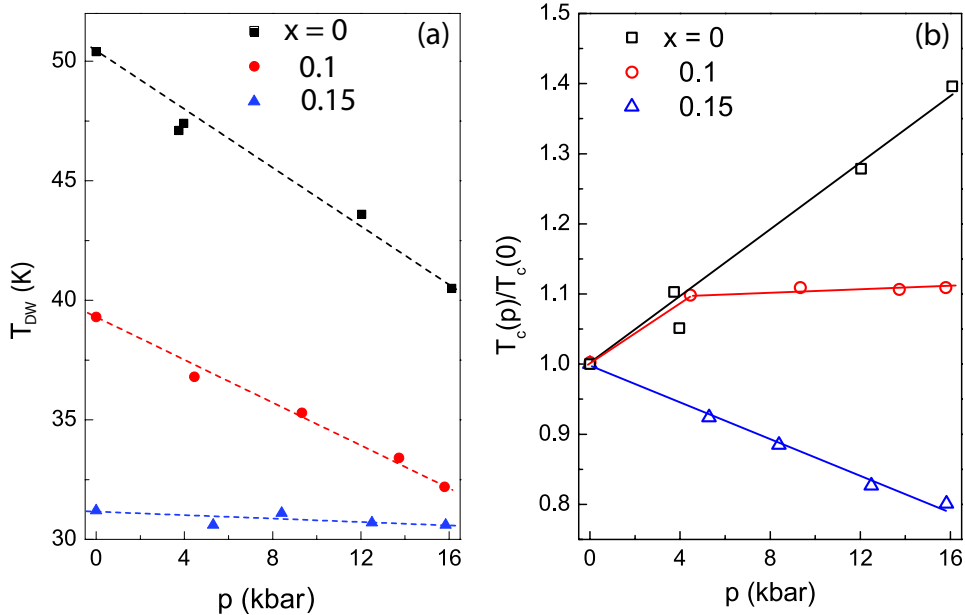,width=5in}}
\vspace*{8pt}
\caption{Pressure effect on superconducting and density wave phases of Ba$_{1-x}$Na$_x$Ti$_2$Sb$_2$O. (a) T$_{DW}$ vs. pressure. (b) T$_c$ vs. pressure. The error bars in (b) show the width of the resistive transition between 90\% and 10\% of the resistance drop.}
\label{f13}
\end{figure}

In superconducting pnictide oxides, the effects of pressure on the superconducting T$_c$ and the density wave order (T$_{DW}$) are of specific interest and have been investigated very recently.\cite{gooch:13b} The hole-doped Ba$_{1-x}$Na$_x$Ti$_2$Sb$_2$O was studied at three typical doping states, for x=0 (undoped), x=0.1 (underdoped), and x=0.15 (optimally doped). The density wave phase was suppressed for all doping levels, however, the rate of decrease of T$_{DW}$ showed a strong dependence on the Na content x, with the largest rate at x=0 and only a very minute decrease of T$_{DW}$ at x=0.15. The pressure effect on the density wave transition temperature is summarized in Fig. \ref{f13}a.

The relative change of the superconducting T$_c$ with pressure is shown in Fig. \ref{f13}b. Similar to most cuprate and iron arsenide superconductors, the pressure coefficient of T$_c$ strongly depends on the doping state. The critical temperature of the undoped compound (BaTi$_2$Sb$_2$O) increases by 40\% with compression to 16 kbar. At 10\% sodium doping, T$_c$ initially increases at low p but then saturates with further increasing pressure. The overall T$_c$ increase does not exceed about 10\%. For the optimally doped sample, however, the opposite trend is observed, the superconducting T$_c$ decreases linearly with increasing pressure.

It should be noted that a similar doping dependence of the pressure coefficient of T$_c$ was reported in high-pressure studies of cuprate high-temperature superconductors\cite{lorenz:05} as well as iron arsenide superconductors.\cite{lorenz:08,gooch:08} In cuprate superconductors, the pressure effects have been understood as a charge transfer to the active superconducting layer induced by pressure and the general dome-shaped T$_c$ vs. doping phase diagram. In iron arsenide superconductors, the proximity of the superconducting state to a spin density wave phase, the possible competition between both phases, and the pressure effect on this phase was considered as an additional circumstance that had to be taken into account.

In Ba$_{1-x}$Na$_x$Ti$_2$Sb$_2$O the active superconducting layer is the Sb-TiO$_2$-Sb slab with charge doping provided by the Na ions occupying Ba sites in between the active layers. A charge transfer induced by pressure to the active layer seems unlikely, but not impossible. However, since the Sb $5p$ states mix with the Ti $3d$ states at the Fermi surface (Sb-Ti bonds), an effect of pressure on the Fermi surface resulting in a redistribution of charges as well as a change of the nesting property cannot be ruled out. In addition, the suppression of the density wave state by pressure, which is particularly pronounced at low doping levels, can contribute to the rise of T$_c$ of the undoped or underdoped compounds. The critical temperature of the density wave state is almost not affected by pressure for optimal doping (Fig. \ref{f13}a) and other effects such as a reduction of the density of states at the Fermi level due to pressure-induced band broadening etc. can dominate and result in the observed T$_c$ decrease for x=0.15.

The apparent anticorrelation between the pressure changes of T$_{DW}$ and T$_c$ lead to the conjecture that both ordered phases actually compete with one another, leading to a T$_c$ enhancement whenever the density wave phase is suppressed by pressure (as in the undoped BaTi$_2$Sb$_2$O). This tendency is similar to the effect of doping on both states, with increasing hole doping T$_{DW}$ decreases and T$_c$ increases. A more fundamental and complete understanding of the effects of external pressure on both states, however, needs additional work, also involving superconducting pnictide oxides with isovalent substitutions.

\subsection{The density wave state: Spin or charge density ?}
\label{DW}
The nature of the density wave phase is crucial for the understanding of superconductivity in pnictide oxides and how both states coexist and accommodate each other. The strong nesting of the Fermi surface, first discussed by Pickett based on first-principles calculations of Na$_2$Ti$_2$Sb$_2$O,\cite{pickett:98} can explain the observed instability (near T$_{DW}$=114 K in Na$_2$Ti$_2$Sb$_2$O) as the formation of a spin or charge density wave with corresponding anomalies of the magnetic susceptibility and the resistivity (see Fig. \ref{f3}). The question of whether the phase below T$_{DW}$ is defined by a spin or a charge density wave has been a matter of discussion since then.

Early studies revealing the sharp drop of the magnetic susceptibility (see Figs. \ref{f3}b, \ref{f4}b, and \ref{f5}b) suggested the opening of a spin gap, possibly due to a spin-Peierls transition.\cite{axtell:97} The high-temperature property of the susceptibility implies that the main magnetic exchange interactions are Ti-Sb-Ti superexchange, but not Ti-O-Ti interactions. Theoretical calculations confirm the hybridization of Ti $d$ states with Sb $p$ states but also emphasize on the importance of direct Ti $d$-$d$ overlap.\cite{pickett:98} Neutron scattering experiments, however, have not detected a sizable magnetic moment that could be associated with static magnetic order.\cite{ozawa:01,nozaki:13,ozawa:00,lynn:12} The absence of static magnetic order is further suggested by muon spin relaxation studies of BaTi$_2$(As$_{1-x}$Sb$_x$)$_2$O and Ba$_{1-x}$Na$_x$Ti$_2$Sb$_2$O.\cite{nozaki:13,rohr:13b}

It should be noted that the absence of a sizable static magnetically ordered moment does not exclude the possibility of the formation of a spin density wave state if the ordered magnetic moment is very small, below the resolution of the neutron or muon spin relaxation experiments. However, the missing proof of a detectable magnetic moment below T$_{DW}$ is in favor of the formation of a charge density wave, with strong coupling of the carriers to the lattice. Anomalies of both lattice constants near T$_{DW}$ have in fact been observed in neutron scattering studies of Na$_2$Ti$_2$Sb$_2$O although no lattice dimerization or superstructures have been reported.\cite{ozawa:00} The structural distortion mainly increases the Ti-Ti distance and reduces the direct overlap with a partial opening of a band gap. The Ti $d^1$ electrons become slightly more localized which explains the increase of the resistivity just below T$_{DW}$. Recent optical spectroscopy experiments show the opening of a density wave-like energy gap at T$_{DW}$ removing most of the free carrier spectral weight with a simultaneous reduction of the scattering rate.\cite{huang:13}

\section{First-principles calculations of the electronic structure}
\label{FPC}
The nature of the density wave order in pnictide oxides (with $Pn$=As, Sb) is one key to understand the superconductivity in these compounds. First theoretical work by Pickett did reveal the nesting property in form of a zone corner (M-point) box-like Fermi surface.\cite{pickett:98} The local spin-density approximation, applied to Na$_2$Ti$_2$Sb$_2$O, confirmed the nearly ionic picture with Na$^{1+}$, Ti$^{3+}$, Sb$^{3-}$, and O$^{2-}$. The valence electrons of Ti$^{3+}$ are in 3$d^1$ state, i.e there is one 3$d$ electron per Ti ion with total spin 1/2. Note the apparent analogy to the 3$d^9$ state of Cu$^{2+}$ in the copper oxide superconductors with one spin-1/2 hole per Cu at the Fermi level. The direct $d$-$d$ orbital overlap and the hybridization with $p$ states from antimony and oxygen result in the delocalization of the Ti $d$ states and the metallic properties of Na$_2$Ti$_2$Sb$_2$O. Accordingly, the density of states at the Fermi energy is mostly composed of Ti $d$ states with some contribution from the Sb and O $p$ orbitals.

The simulation of different kinds of ferromagnetic or antiferromagnetic long range order by Pickett\cite{pickett:98} have not revealed a stable magnetic ground state with ordering of the magnetic Ti 3$d^1$ moments. Therefore, the formation of a spin- or charge-density wave which originates from the nesting property of the box-like section of the Fermi surface with mostly Ti $d$ character was suggested.

The existence of nested sections of the Fermi surface in NaTi$_2$Sb$_2$O around the M point with a pseudo one-dimensional character was confirmed by a tight-binding band structure calculation (extended H\"{u}ckel method).\cite{biani:98} It was suggested that a charge density wave with an incommensurate modulation and an associated lattice distortion would lead to the anomaly of the resistivity observed at T$_{DW}$=120 K for this compound (Fig. \ref{f3}b). The structural distortion affecting the $a$ and $c$ axis lattice parameters and the $a/c$ ratio was indeed observed in powder neutron scattering experiments,\cite{ozawa:00} but no superlattice reflections of magnetic or lattice origin could be resolved.

For a better understanding of the electronic structure, the character of the occupied Ti $d$ bands is essential. The nested box Fermi surface is comprised of almost exclusively Ti $d_{xy}$ states with a significant overlap between the orbitals of neighboring Ti ions.\cite{pickett:98,biani:98} The Ti orbitals hybridize with the Sb $p$ bands closer to the center of the Brillouin zone. It should be noted that this is different from the Cu $3d$ orbitals in the copper oxide superconductors which have prominently $d_{x^2-y^2}$ character and mix with the oxygen $p$ bands.

\begin{figure}[bt]
\centerline{\psfig{file=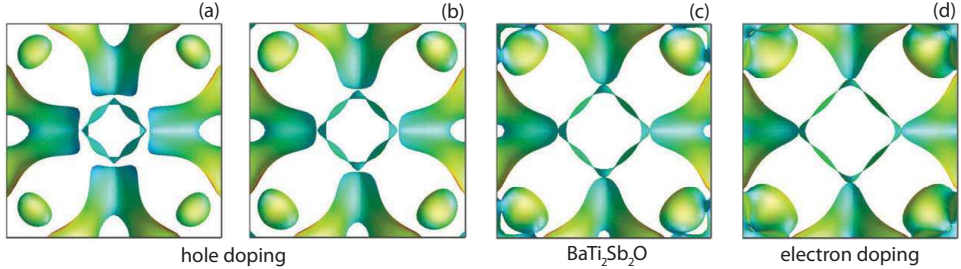,width=5in}}
\vspace*{8pt}
\caption{Fermi surface of doped BaTi$_2$Sb$_2$O viewed along the $c$-axis. (a) and (b): $E_F<0$, hole doping (Ba substituted by Na, K, Rb), (c): $E_F=0$, BaTi$_2$Sb$_2$O, (d): $E_F>0$, electron doping. Reprinted with the publisher's permission from D. J. Singh, NJP \textbf{14}, 123003 (2012).}
\label{f14}
\end{figure}

The problem of possible magnetic order and its consequences for superconductivity in Ba$_{1-x}$Na$_x$Ti$_2$Sb$_2$O was discussed by Singh based on first-principles density functional calculations.\cite{singh:12} Similar to Na$_2$Ti$_2$Sb$_2$O, discussed in the previous paragraphs, the nearly ionic picture (Ba$^{2+}$, Ti$^{3+}$, Sb$^{3-}$, O$^{2-}$) holds very well with Ti $d$ states hybridized with Sb $p$ states providing most of the density of states near the Fermy energy. The Ti $d$ orbitals at the Fermi surface are a mixture of $d_{z^2}$, $d_{x^2-y^2}$, and $d_{xy}$, with a strong $d_{xy}$ character. The Fermi surface obtained is rather complex with electron- as well as hole-like sheets present. Most importantly, there exists a 2D square section stretching along the zone edge line (M-A) giving rise to the nesting property. Another nested region of the Fermi surface extends around the X point. The nesting wave vector along the orientation of the Ti square lattice is (0.24,0.24,0)(2$\pi$/a), in good agreement with earlier calculations.\cite{biani:98} The calculated Fermi surface is shown in Fig. \ref{f14} for different positions of the Fermi energy (corresponding to different doping levels). The nesting property of the box-shaped section around M-A is remarkably stable with respect to a variation of the Fermi energy $E_F$. Decreasing $E_F$ ($E_F$ is set to zero for BaTi$_2$Sb$_2$O) below zero, corresponding to hole doping with Na, K, or Rb, decreases the size but the square shape is approximately preserved. This may explain the extraordinary stability of the density wave phase upon hole doping (see Fig. \ref{f10}), as first observed experimentally in Ba$_{1-x}$Na$_x$Ti$_2$Sb$_2$O.\cite{doan:12}

A magnetic instability was found (due to the Fermi surface nesting) resulting in an ordered ground state with a double stripe spin structure which also lowers the symmetry from tetragonal to orthorhombic.\cite{singh:12} The ordered magnetic moment, however, is small ($\simeq 0.2\mu_B$), according to the calculations. Nevertheless, the coupling of the spin to the charge carriers appears to be significant with possible consequences for the superconducting state. The possibility of spin-fluctuation mediated superconductivity was consequently discussed and the pairing symmetry, consistent with the structure of the Fermi surface, was suggested to be a sign changing $s$-wave superconducting state. For this "unconventional" superconductivity it is important that the spin fluctuation dominate and the transition at T$_{DW}$ results in the formation of a spin density wave.

More recent first-principles calculations of Na$_2$Ti$_2Pn_2$O ($Pn$=As, Sb) have also concluded that the ground state is magnetically ordered.\cite{yan:13} The calculated band structures and Fermi surfaces are consistent with earlier calculations.\cite{pickett:98} The magnetic structure of the ground state depends on the pnictogen atom. For $Pn$=As, a checkerboard spin configuration was found to be stable, with four spins on a plaquette of neighboring Ti ions aligned with one another, forming a block spin with an antiferromagnetic order between different blocks. For $Pn$=Sb the magnetic order is the double stripe structure described by Singh for BaTi$_2$Sb$_2$O. The Fermi surface nesting as the origin of the magnetic orders is confirmed. A slightly different picture of the magnetic ground state was derived from electronic structure calculations of BaTi$_2$As$_2$O.\cite{yu:14} Only two of four Ti spins were found to order with antiparallel moments, driven by the Fermi surface nesting and the Coulomb interactions. No modulation of the charge density was found, favoring the formation of a spin density wave at the ordering temperature T$_{DW}$.

It appears interesting that first-principles calculations of various pnictide-oxides reveal different ground states with double stripe, block spin, or partial magnetic orders. These magnetic structures are very close in energy, but they are all commensurate with the lattice. The nesting wave vector, however, is most likely incommensurate with the atomic structure and the true magnetic order may well be an incommensurate spin density wave. With the small ordered moment per Ti ion, it could be a challenge to detect the possible order in neutron scattering or other experiments.

First-principles calculations including the phonon dispersion spectrum and the electron-phonon interactions of BaTi$_2$Sb$_2$O have been presented by Subedi.\cite{subedi:13} It was concluded that, despite the tendency to a magnetic instability, the phonon dispersions may give rise to a lattice instability near the corners of the Brillouin zone which stabilizes a charge density wave phase. The electron-phonon coupling is strong enough to explain the superconducting phase in this compound as conventional phonon-mediated s-wave superconductivity. The calculated electron-phonon coupling constant, $\lambda_{ep}$=0.55, is in good agreement with the experimentally derived value.\cite{gooch:13}

An analysis of the electronic structure and chemical bonding of Na$_2$Ti$_2Pn_2$O and BaTi$_2Pn_2$O ($Pn$=As, Sb) has in part confirmed the results from earlier work.\cite{suetin:13} The bonding between the Ti$_2Pn_2$O and Na$_2$ (Ba) layers is mainly of ionic character, whereas the bonding within the Ti$_2Pn_2$O has also covalent character. Ti $d_{xy}$, $d_{x^2-y^2}$, and $d_{z^2}$ states dominate the Fermi surface. Interestingly, the charge transfer to the Ti$_2Pn_2$O layer is smaller than expected from the nominal charges Na$^{1+}$ and Ba$^{2+}$. It decreases from 1.6 electrons per Na$_2$ in Na$_2$Ti$_2$As$_2$O to 1.23 electrons per Ba in BaTi$_2$Sb$_2$O.\cite{suetin:13} This trend in the charge transfer from the cation to the Ti$_2Pn_2$O layer could be one relevant parameter which controls the superconducting state in pnictide oxides. It is interesting that superconductivity occurs among the four compounds only in BaTi$_2$Sb$_2$O (T$_c$=1.2 K) with the smallest electron transfer and T$_c$ was enhanced by Na (hole) doping, i.e. by a further reduction of the electron number in the active layer. Similar calculations for BaTi$_2$Bi$_2$O, a pnictide oxide with no density wave phase but a relatively high superconducting T$_c$=4.6 K, show an even smaller charge transfer of 1.18 electrons per Ba ion,\cite{suetin:13b} which is consistent with the general trend mentioned above.

The role of the charge density in the active Ti$_2Pn_2$O layer is not completely understood and needs further studies. Within the rigid band model, the decrease of the electron number will change the density of states (DOS) at the Fermi energy, which is expected to increase since the Fermi energy is just on the high-energy slope of a peak of the DOS. On the other hand, the superconducting state appears to compete with the density wave state and reducing T$_{DW}$ or suppressing the DW state completely leads to a superconducting phase, observed in BaTi$_2$Sb$_2$O, its hole-doped analogues, and BaTi$_2$Bi$_2$O. Further experimental and theoretical studies are needed to deepen the understanding of the coexistence and mutual interactions of the density wave and superconducting states in the new Ti-based pnictide oxide superconductors.

\section{Summary}
\label{S}
Superconductivity in a novel transition metal pnictide oxide has attracted attention recently. This class of compounds with the chemical compositions Na$_2$Ti$_2Pn_2$O and BaTi$_2Pn_2$O ($Pn$=As, Sb, Bi) form layered structures with characteristic Ti$_2Pn_2$O slabs separated by either a double layer of Na$_2$ or a single layer of Ba. Here the transition metal is Ti with one 3$d$ electron in the formal ionic Ti$^{3+}$ state. Structurally, the Ti$_2$O form a 2D lattice which is the anti-structure to the CuO$_2$ planes of the copper oxide high temperature superconductors, i.e. the titanium and oxygen ions occupy the oxygen and copper sites in the CuO$_2$ structure, respectively. The pnictogen ions are located above and below the Ti$_2$O plane.

With the exception of BaTi$_2$Bi$_2$O, all other compounds exhibit an instability at higher temperature which is most probably caused by a nesting feature of sections of the Fermi surface. The nature of this instability is not completely resolved yet, but there is evidence for the formation of a density wave state, through a modulation of either charge or spin density. The electronic states at the Fermi energy are dominated by Ti $d$ states partially hybridized with $p$ states of $Pn$ and O.

Superconductivity was discovered in BaTi$_2$Sb$_2$O at 1.2 K and T$_c$ did increase through Na doping to a maximum of 5.5 K. Interestingly, the density wave state was only slightly suppressed by Na (hole) doping and it coexists with the superconducting state in Ba$_{1-x}$Na$_x$Ti$_2$Sb$_2$O up to x=0.35, the solubility limit of Na in the structure of BaTi$_2$Sb$_2$O. Similar results have been obtained for other hole-doped samples using K and Rb replacing Ba, or Sn replacing Sb. The nature of the superconducting pairing has been a matter of discussion. It may be closely related to the nature of the density wave phase, spin or charge density. Various experimental techniques, designed to reveal information about the symmetry of the superconducting order parameter, have provided evidence in favor of a conventional BCS-like $s$-wave superconductivity. Furthermore, no signature of a magnetically ordered state could be detected although first-principles calculations did find stable magnetic ground states in different pnictide oxide compounds.

Superconductivity and the density wave order most likely compete with one another. Evidence for this competition is derived from the fact that a superconducting state is only observed when the critical temperature of the density wave transition is lowered to 54 K (in BaTi$_2$Sb$_2$O) or completely suppressed (in BaTi$_2$Bi$_2$O). Furthermore, doping and pressure dependent measurements show an anticorrelation between T$_c$ and T$_{DW}$, T$_c$ is enhanced when T$_{DW}$ decreases.

On a microscopic level, there are several parameters that may affect the two ordered phases. With the increase of the $Pn$ ionic size, the $a$- and $c$-axis lattice parameters increase in both, Na$_2$Ti$_2Pn_2$O and in BaTi$_2Pn_2$O. Comparing Na$_2$Ti$_2Pn_2$O and BaTi$_2Pn_2$O, $a$ and $c$ are slightly smaller in the Ba compounds. An increase of $a$ will reduce the direct overlap of the Ti $d$ orbitals and also affect the hybridization with the $Pn$ $p$ states. Furthermore, the calculated charge transfer from the (Na$^{1+}$)$_2$ and the Ba$^{2+}$ layers is not complete and changes with the cation (Na or Ba) and with $Pn$. It turns out that the smallest number of electrons is transferred to the Ti$_2Pn_2$O layer in BaTi$_2$Bi$_2$O, the compound with no density wave phase and the highest superconducting T$_c$.

There are a number of questions that should be addressed in future studies of pnictide oxide superconductors. The experimental verification and the theoretical understanding of the nature of the density wave phase is crucial to understand the competition and mutual interaction of this phase with the superconducting state. The solid solution BaTi$_2$(Sb$_{1-x}$Bi$_x$)$_2$O shows superconductivity only for small and large values of x. The reason for the missing superconducting state around x=0.5 has yet to be explored. The effects of electron doping, if achievable, on the density wave and superconducting states is of interest since it could help to understand the role of the electron density in the Ti$_2Pn_2$O layers. The Ti-based pnictide oxides are an interesting class of layered compounds and future work will certainly contribute to a deeper understanding of superconductivity and density wave instabilities and how those states interact with one another.

\section*{Acknowledgments}
The authors gratefully acknowledge support from the T.L.L. Temple Foundation, the John J. and Rebecca Moores Endowment, the State of Texas through the Texas Center for Superconductivity at the University of Houston, the US Air Force Office of Scientific Research (FA 9550-09-1-0656), the National Science Foundation (CHE-0616805), and the R. A. Welch Foundation (E-1297).

\section*{References}

\bibliographystyle{prsty}

\end{document}